# Synthesis of WTe$_2$ thin films and highly-crystalline nanobelts from pre-deposited reactants

*John B. Mc Manus*[1,2], *Cansu Ilhan*[1,2], *Bastien Balsamo*[2,3], *Clive Downing*[1,2], *Conor P. Cullen*[1,2], *Tanja Stimpfel-Lidner*[4], *Graeme Cunningham*[1,2], *Lisanne Peters*[1,2], *Lewys Jones*[2,5], *Daragh Mullarkey*[5], *Igor V. Shvets*[5], *Georg S. Duesberg*[1,4], *Niall McEvoy*[1,2]\*

[1] School of Chemistry, Trinity College Dublin, Dublin 2, D02 PN40, Ireland

[2] AMBER Centre, CRANN Institute, Trinity College Dublin, Dublin 2, D02 PN40, Ireland

[3] SIGMA Clermont, Université Clermont Auvergne, F-63000 Clermont–Ferrand, France

[4] Institute of Physics, EIT 2, Faculty of Electrical Engineering and Information Technology, Universität der Bundeswehr, 85579 Neubiberg, Germany

[5] School of Physics, Trinity College Dublin, Dublin 2, D02 PN40, Ireland

\*Corresponding author: nmcevoy@tcd.ie

ABSTRACT

Tungsten ditelluride is a layered transition metal dichalcogenide (TMD) that has attracted increasing research interest in recent years. $WTe_2$ has demonstrated large non-saturating magnetoresistance, potential for spintronic applications and promise as a type-II Weyl semimetal. The majority of works on $WTe_2$ have relied on mechanically-exfoliated flakes from chemical vapour transport (CVT) grown crystals for their investigations. While producing high-quality samples, this method is hindered by several disadvantages including long synthesis times, high-temperature anneals and an inherent lack of scalability.

In this work, a synthesis method is demonstrated that allows the production of large-area polycrystalline films of $WTe_2$. This is achieved by the reaction of pre-deposited films of W and Te at a relatively low temperature of 550 ˚C. Sputter X-ray photoelectron spectroscopy reveals the rapid but self-limiting nature of the oxidation of these $WTe_2$ films in ambient conditions. The $WTe_2$ films are composed of areas of micrometre sized nanobelts that can be isolated and offer potential as an alternative to CVT-grown samples. These nanobelts are highly crystalline with low defect densities indicated by TEM and show promising initial electrical results.

1. INTRODUCTION

Transition metal dichalcogenides (TMDs) are a family of layered materials whose nanoscale forms have been extensively studied over the last number of years. This interest is due to their varied and layer-dependent properties, which gives them a wide range of potential applications.(1, 2)

Unlike some of the more commonly-studied TMDs, such as $MoS_2$ and $WS_2$ that exist predominantly in the trigonal prismatic 2H phase, $WTe_2$ is most stable in the distorted octahedral $T_d$ phase, making it unique among the Group VI TMDs.(3, 4) This structure means WTe$_2$ is semimetallic and opens up unique electronic properties, such as topological electronic states.(5) Thorough investigations of the structure and electrical properties of bulk WTe$_2$ were carried out in the 1950's and 1960's, with particular emphasis placed on its thermoelectric properties.(3, 6-11) However, it is only recently that investigations have focussed on mono- and few-layer forms.

WTe$_2$ can be considered a layered or 2D material, though it does also have some 1D character due to its distorted octahedral structure. This distortion causes quasi-1D chains of W atoms to be formed within the layers, schematically shown in Fig.1 (a) insert.(4) This leads to a strong anisotropy in properties such as the conductivity and mechanical response within the monolayer.(12) It also manifests in the tendency of WTe$_2$ crystals to grow anisotropically.(13) The crystals are generally longest along the b crystallographic direction, parallel to the 1D tungsten chains.

Applications-centred studies of WTe$_2$ have historically focussed on thermoelectrics but more recent studies have examined a much broader range of potentially interesting properties and applications. Experimental investigations of the electronic properties of WTe$_2$ have demonstrated large, non-saturating magnetoresistance and high charge-carrier mobilities of up to 10,000 cm$^2$V$^{-1}$s$^{-1}$.(14-16) There have also been reports of ferroelectricity(17) and superconductivity(18, 19) in WTe$_2$. Furthermore, it was the first material to be suggested as a potential type-II Weyl semimetal.(20) Since this prediction, there have been a number of experimental reports that have lent credence to this assertion.(21, 22)

Other works have examined WTe$_2$ for use in applications such as nanoscale electrical interconnects(23), electrocatalysis(24, 25) and as an anode material for Na-ion batteries.(26) Additionally, WTe$_2$ has been touted as a promising candidate for large-gap quantum spin Hall

insulators(5, 21, 27) Lastly, through electrochemical investigations the possibility of using $WTe_2$ as a catalyst for the hydrogen evolution reaction (HER) has been examined.(24, 28-31)

The majority of reports, especially those examining its electronic properties, have focused on mechanically-exfoliated flakes from chemical vapour transport (CVT) grown crystals of $WTe_2$.(14-16, 32-34) While CVT does produce exceptionally high-quality crystals, it is a time and energy intensive process requiring long anneals at temperatures of up to 1000 ˚C. The subsequent laborious mechanical exfoliation and transfer process also limits the scalability of any devices made using this method.

There has been some work on the bottom-up growth of nanoscale $WTe_2$, however large-area, reproducible synthesis remains an open challenge.(35, 36) Table 1 gives an overview of a range of publications that synthesised $WTe_2$ on the nanoscale. Chemical vapour deposition (CVD) growth of $WTe_2$ has achieved single-crystal monolayers with lateral dimensions on the order or 10s to 100s of micrometres from Te, $WO_3$ and $WCl_6$ precursors.(35-37) Despite requiring high growth temperatures, 650 – 800 ˚C, CVD growth remains the most promising method for high-quality monolayer synthesis. Nevertheless, there remains significant challenges to overcome before large-area, reproducible, layer controlled growth of $WTe_2$ is achieved.

Large-area growth of polycrystalline films of $WTe_2$ has however been achieved, with thicknesses ranging from a few to tens of monolayers.(38-41) These films are generally synthesised by the conversion of a pre-deposited W source on a substrate. This is accomplished by exposure of the W source to Te vapour in a controlled atmosphere at an elevated temperature. These methods offer the advantage of producing large-area films in a scalable manner. The final thickness of the film can also be determined by the thickness of the initially deposited W source. These polycrystalline films have grain sizes ranging from less than 10 nm to ~100 nm. This means that while promising for a range of applications, these films are not suitable for many electronic applications due to high and variable levels of unintentional doping and high densities of scattering centres.(42)

A number of other works have synthesised $WTe_2$ on the nanoscale using methods such as solution-based growth or molecular beam epitaxy (MBE).(26, 43, 44)

This discussion shows that while substantial progress has been made in the synthesis of nanoscale $WTe_2$, the field is far from mature. The majority of works rely on CVT meaning there is a pressing need for further work on alternative lower temperature synthesis methods, especially those which are scalable.

Table 1: Literature review of published synthesis techniques to produce nanoscale $WTe_2$.

| Ref. | Method Description | Subsequent processing to obtain 2D form | Material Characteristics | Synthesis T (°C) | Growth Time |
|---|---|---|---|---|---|
| (14) | CVT | Mechanical exfoliation(ME) | Single-crystal flakes | 750 | 11 days |
| (45) | CVT | ME | Single-crystal flakes | 800 | 10 days |
| (32) | CVT | ME | Single-crystal flakes | 900 | 4 days |
| (46) | CVT | ME | Single-crystal flakes | 1000 | 8 days |
| (22) | CVT | ME | Single-crystal flakes | 800 | 11 days |
| (47) | CVT | ME | Single-crystal flakes | 1020 | 5 days |
| (48) | CVT | ME | Single-crystal flakes | 950 | 3 days |
| (33) | CVT | ME | Single-crystal flakes | 1100 | 4 days |
| (35) | CVD, ammonium tungstate hydrate and KCl promoter | None | Monolayer flakes ~350 µm | 800 | 20 min |
| (37) | CVD, $WCL_6/WO_3$ precursor | None | Monolayer flakes ~350 µm | 820 | 20 min |
| (38) | CVD $WCl_6$ precursor | None | Polycrystalline film; thickness above 5 nm | 500 | 20 min |
| (36) | CVD, ammonium metatungstate precursor | None | Monolayer flakes ~50 µm | 650 | 6 min |
| (49) | CVD, $WO_3$ precursor and KI growth promoter | None | Monolayer flakes 10s µm, | 700 | 1 hour |
| (39) | Film conversion, predeposited $W/WO_3$ | None | Large-area polycrystalline film, thickness 6 nm, grain size ~6 nm | 650 | 60 min |
| (50) | Film conversion, predeposited $WO_3$ nanowires | None | Nanowires | 500 | 10 min |
| (51) | Film conversion, predeposited W, close proximity $Ni_xTe_y$ source | None | Large-area polycrystalline films, thickness - few layer to bulk, 10s nm grain size | 500 | 10-30 min |
| (40) | Film conversion, predeposited W, need $H_2$ present during reaction | None | Large-area polycrystalline film, thickness 1 + nm, grain size 10s nm | 800 | 50 min |
| (41) | Film conversion, predeposited W and Cu | None/ Mechanical transfer | Large-area polycrystalline film of $WTe_2$ nanobelts | 500 | 10 min |
| (44) | MBE, HOPG/$MoS_2$ substrates | None | Large-area films | 275 | 1.5nm hr$^{-1}$ |
| (52) | CVT | Sonication | Quantum dots | | |
| (53) | Pulsed laser deposition | None | Polycrystalline film; thickness 5+ nm | | |
| (54) | Pulsed laser deposition | Post-deposition anneal with Te | Highly crystalline, thickness ~100 nm | 700 | 1 nm min$^{-1}$ 48hr anneal |
| (25) | Solution-phase growth | | Nanostars | | |
| (55) | Solution-phase growth | | Nanostructures, ~200 nm | 260 | |
| (43) | Solution-phase film formation, $WCl_4$ precursor | Microwave heating | Large-area polycrystalline film, few-layer thickness, grain size 6 nm | | |

In this work we present a method to synthesise polycrystalline films consisting of $WTe_2$ nanobelts, with individual nanobelts of up to 10 µm in length. These are synthesised by annealing pre-deposited films of W and Te at a relatively low temperature of 550 ˚C. Characterisation of the films reveals that their surface is prone to oxidation but this is self-limiting and does not extend into the bulk. Due to their large lateral size, these micrometre-scale, single-crystal nanobelts of $WTe_2$ can be isolated and characterised individually. The high crystallinity of the nanobelts implies their suitability for applications previously demonstrated using CVT-grown flakes. To show this, individual nanobelts were electrically characterised. The ability to access both large-area films and single-crystalline flakes demonstrates the versatility of this method to produce high-quality samples of $WTe_2$ in an efficient manner.

## 2. RESULTS AND DISCUSSION

### 2.1 Synthesis of WTe$_2$ films

Films of WTe$_2$ were synthesised by depositing a layer of W onto a target substrate using Ar-ion sputtering, followed by the electrodeposition of Te on top of this, as shown in Fig. 1(a). The thickness and pattern of the W layer was defined during the sputtering process, while the electrodeposition allowed the Te to be deposited in a controlled and directed manner onto each sample. This method gave local, well-defined quantities of both reactants directly on the sample. This avoided the need for a remote Te source and the challenges associated with this, such as requiring high excess of reactant to be used.

The layers of W and Te were then placed inside nested crucibles and then loaded into a quartz furnace, where they were heated to form WTe$_2$, shown in Fig. S1(a) The conversion typically took place at a dwell temperature of 550 °C for a period of 90 minutes under an inert atmosphere.

All samples are named based on their starting W thickness, as measured by the deposition tool's quartz crystal monitor (QCM). The sample substrates were SiO$_2$/Si wafer with a ~300 nm layer of pyrolytic carbon (PyC) to improve the uniformity of the Te film deposition.(56) The samples studied were ~1 x 1 cm in area but the process could be scaled up to produce larger area samples.

This synthesis method allowed the growth of large-area polycrystalline films of WTe$_2$. Fig. 1(a)-(c) shows scanning electron microscopy (SEM) images of a typical WTe$_2$ film synthesised from an initial 20 nm W layer. It is immediately apparent that the surface of the film consists of two differing morphologies. There are discrete patches, with dimensions of up to tens of micrometres, containing elongated WTe$_2$ nanobelts. The nanobelts are several micrometres in length and have widths of tens to hundreds of nanometres. Surrounding the patches of nanobelts are areas of smaller grains with lengths below 500 nm. The growth of elongated nanobelts is due to the quasi-1D nature of WTe$_2$, similar growth of elongated WTe$_2$ flakes that has been seen in CVD growth of WTe$_2$ in a number of works.(26, 35, 57) The 1D chains of W in the lattice would be expected to be parallel to the long axis of the nanobelts.

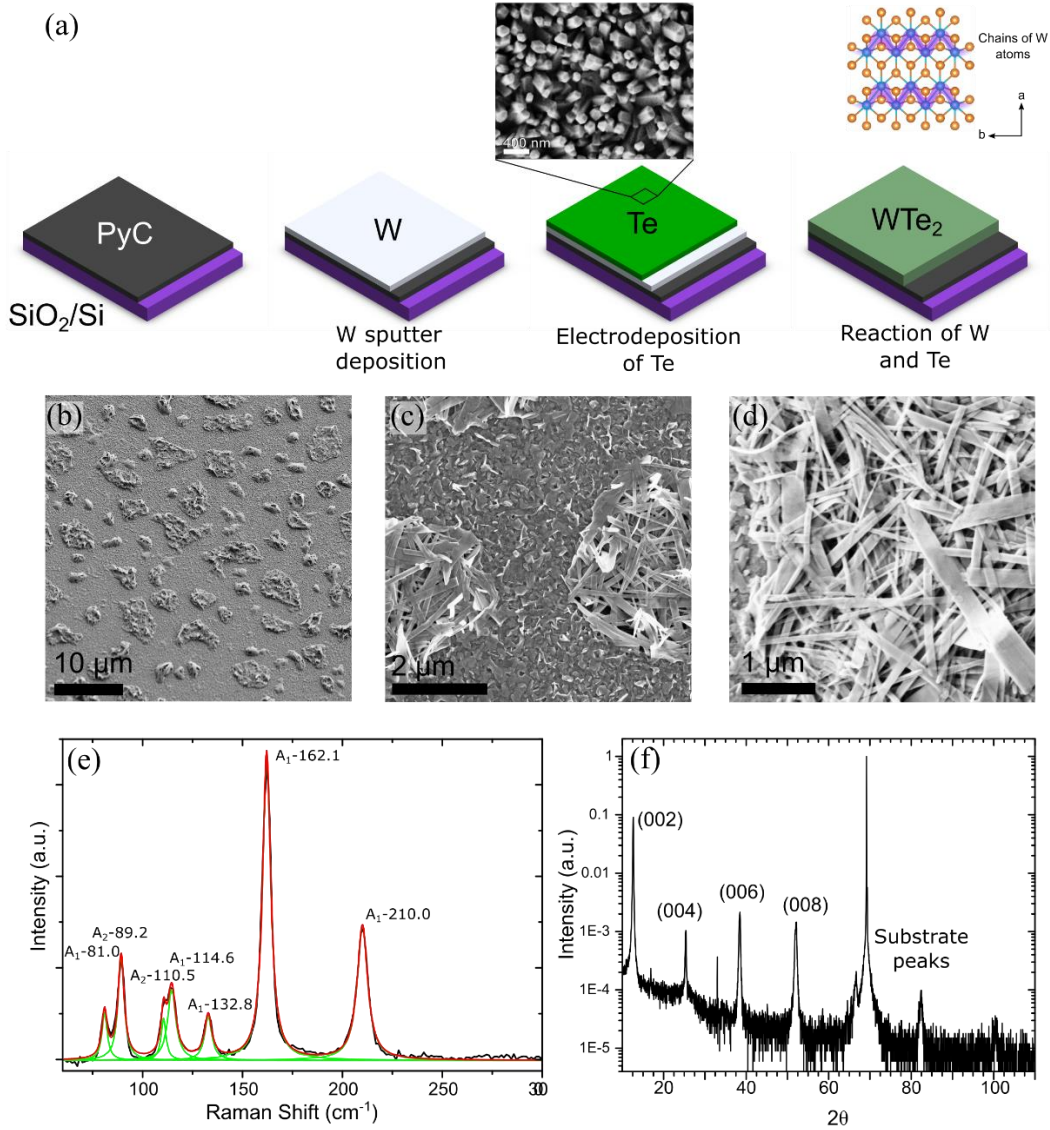

**Fig.1** (a) The process steps involved in WTe$_2$ film synthesis with an SEM image of the surface of the TE film before conversion to WTe$_2$. Insert: Diagram of the 1D W atom chains in the WTe$_2$ structure. (b)-(d) SEM images of the surface of WTe$_2$ film synthesised from 20 nm initial W layer. (d) Raman spectrum of the same film showing the expected modes of bulk-like WTe$_2$. Modes are fitted with Lorentzian peaks to extract positions. (e) XRD diffraction pattern of WTe$_2$ film showing the (002) and associated reflections. A number of substrate peaks are also visible above 60 degrees.

The WTe$_2$ films are synthesised in a straight-forward, scalable manner with full coverage over the entirety of the sample. This compares favourably with the majority of previously-reported growth methods for WTe$_2$ which are by-in-large laborious, time consuming and not scalable. Furthermore, the properties of the films – thickness, grain size – can be tailored by changing the parameters of the initial W and Te depositions.

Compared to typical film conversion methods, which have grain sizes of 10s of nanometres, this work allows the growth of films made up of much larger crystals, on the order of micrometres. This is likely due to the growth occurring at a solid-liquid interface, rather than the typical solid-vapour one. Having the liquid Te in direct contact with the W layer counteracts the low activity of Te and facilitates the synthesis of WTe$_2$ at the comparatively low synthesis temperature and short time, similar to other works using Ni$_x$Te$_y$ alloys.(51) Furthermore, H$_2$ is not required to enable the growth in this work, unlike previously reported methods.(40)

The morphology of these WTe$_2$ films is also influenced by the material properties - MoTe$_2$ and PtTe$_2$ previously grown using this method were found to have significantly different morphologies, with neither showing films consisting of large nanobelts.(56, 58)

Raman analysis, shown in Fig 1(d), provides confirmation of the formation of WTe$_2$. This spectrum closely matches previously published spectra of bulk-like WTe$_2$, indicating the successful growth of T$_d$ phase WTe$_2$ using this method.(59-61) The peaks at 81.0 cm$^{-1}$, 89.2 cm$^{-1}$, 114.6 cm$^{-1}$, 132.8 cm$^{-1}$, 162.1 cm$^{-1}$ and 210.0 cm$^{-1}$ are assigned the symmetry A$_1$, while the peak at 110.5 cm$^{-1}$ is A$_2$.(32, 62) The Raman spectrum does not indicate any modes associated with oxides or other contaminants in the films.

Confirmation of the presence of WTe$_2$ throughout the surface, was provided by Raman spectroscopy mapping, Fig. S2. While there are two morphologies present on the surface, which look quite different under optical microscopy and SEM, Raman spectroscopy mapping confirms that they are both $T_d$ WTe$_2$.

X-ray diffraction (XRD) further confirmed the conversion of the W and Te films to WTe$_2$. XRD results are shown in Fig. 1(e), the diffraction pattern measured is consistent with previously-reported XRD of WTe$_2$ with the (002), (004), (006) and (008) peaks being most prominent.(3) As the film is polycrystalline, there is a significant background and noise in the diffraction pattern. This made further analysis, such as extracting average grain size, infeasible.

To obtain more quantitative and stoichiometric information, the films were analysed using X-ray photoelectron spectroscopy (XPS). This also gave information on the stability of the films, which can be considered important as previous reports have indicated a strong tendency for WTe$_2$ to oxidise in ambient conditions. (36, 48, 63)

The XPS results are shown in Fig. 2. Shown in (a) is the W 4f core-level region of WTe$_2$, the W 5p and Te 4d core levels are also visible in this energy window however the analysis here will only focus on the W 4f core levels, as these are the most intense. There are two doublets present for the W 4f core level, one associated with WTe$_2$ and the other with WO$_3$. Of the W atoms, 30% are in the form of WO$_3$, with the remaining 70% in WTe$_2$. Similarly in Fig.2 (b), there are two doublets associated with the Te 3d core level, one corresponding to Te bound to W as WTe$_2$ and the other for Te in the form of oxides. The Te on the surface is predominantly in the form of WTe$_2$, with only 8% of the atoms bound to oxygen. The stoichiometry of the WTe$_2$ on the surface was found to be WTe$_{1.8}$, indicating a slight deficiency in Te. Finally, the asymmetric nature of the WTe$_2$ peaks indicates the (semi-)metallic nature of the sample.

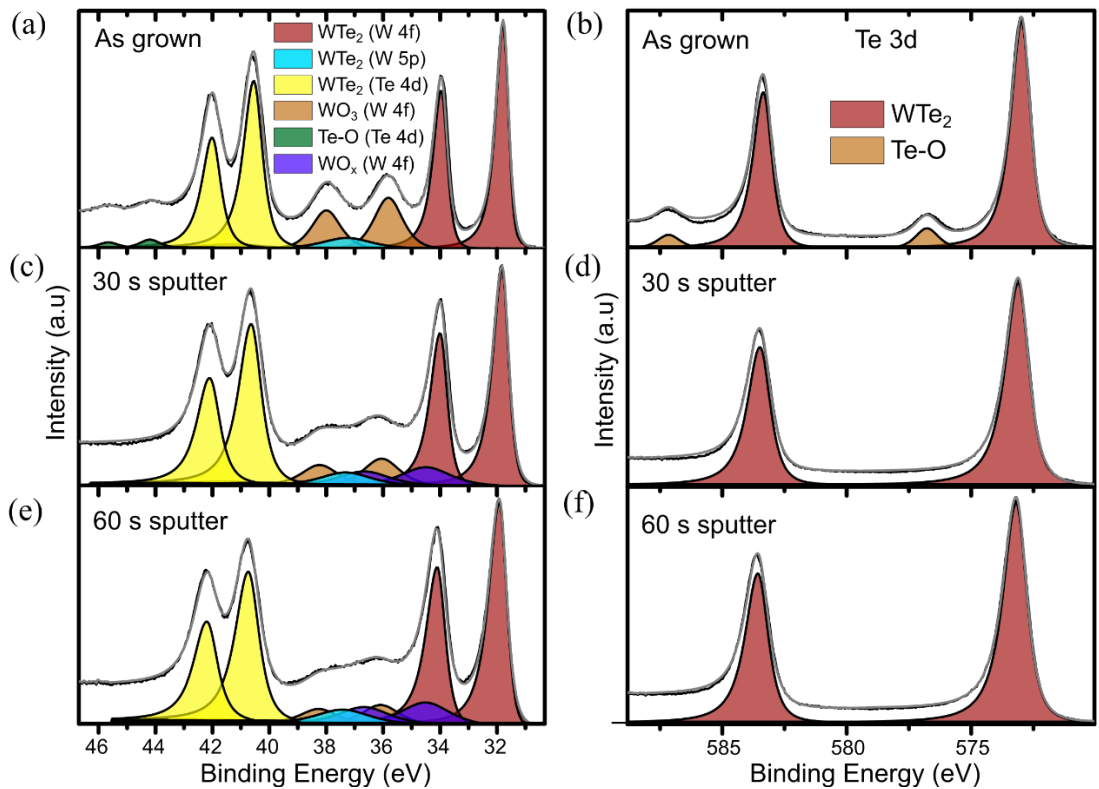

**Fig.2** (a), (c), (e) XPS spectra of the W 4f region for a WTe$_2$ sample as grown, after 30 s of sputtering and after 60 s of sputtering respectively. (b), (d), (f) The same data for the Te 3d binding energy region of the WTe$_2$ sample

## 2.2 Oxidation of WTe$_2$ films

The level of oxidation observed on the surface of the WTe$_2$ film implies that it is prone to degradation in atmosphere, more so than other telluride TMDs such as MoTe$_2$ and PtTe$_2$ that have shown lower oxidation levels when similarly analysed.(56, 58) Previous works on the susceptibility of WTe$_2$ to oxidation have concluded that the oxidation of the first few atomic layers of WTe$_2$ is extremely rapid, however the oxidation is self-limiting and tends not to extend far into the bulk. (36, 48, 63) It is this surface layer (less than 10 nm) that is exclusively probed by XPS, potentially leading to an over estimation of the over-all oxidation level of the sample. To offer further understanding of this, the sample was analysed using sputter XPS.

Sputter XPS involves repeated cycles of in-situ sputtering and XPS analysis, allowing the elemental composition at both the surface and at various depths in the bulk of the sample to be measured. The sputter XPS process is described in more detail in the methods section.

Two rounds of 30 seconds Ar sputtering were carried out, with the results shown in Fig. 2(c) – (f). The Te 3d core-level spectra show that after 30 seconds of sputtering only a single doublet corresponding to Te in the form of $WTe_2$ is visible. The signal of Te bound to oxygen is below the detection limit of the tool, indicating that the Te oxide is only present on the surface with virtually none in the bulk of the sample.

The signal from $WO_3$ reduces significantly after sputtering, from about 30% initially, to 10% after 60 seconds. During the same time, the percentage of W atoms in the form of $WTe_2$ remains roughly constant at 70%. The difference is accounted for by the development of a $WO_x$ signal that accounts for slightly less than 20% of W atoms after 60 seconds of sputtering. This $WO_x$ peak is not present in the surface scan but develops after sputtering. All the W peaks broaden after sputtering, due to damage or amorphization caused by the sputtering process.

The sputtering XPS shows that $TeO_2$ and $WO_3$ are primarily present on the surface of the films. However, a $WO_x$ signal is present after sputtering, and the percentage of W atoms present as either $WO_3$ or $WO_x$ is roughly constant throughout. Furthermore, the overall ratio of W to Te on the surface changes after sputtering, with the sample becoming increasingly deficient in Te. The proposed explanation of these observations is that the Te is being preferentially sputtered, meaning that the sample becomes more W rich during this process, and the $WO_3$ is not being effectively sputtered but rather becoming a substoichiometric $WO_x$.(64) This also correlates with previous reports that indicate the self-limiting nature of $WTe_2$ oxidation. Furthermore, this conclusion also aligns with why no strong oxide signal is observed in the Raman spectroscopy – Raman is much less surface-sensitive than XPS so the WTe$_2$ signal remains dominant.

## 2.3 WTe$_2$ growth mechanism

The majority of films in this work were synthesised at a temperature of 550 ˚C, however WTe$_2$ was successfully grown over a range of temperatures. Fig. 3(a) shows Raman spectra of films synthesised at a range of temperatures. At 350 ˚C, the only Raman peaks visible are attributed to elemental Te.(65) For all growth temperatures between 450 ˚C and 650 ˚C characteristic peaks associated with bulk-like WTe$_2$ are observed. No discernible peaks are visible in the 750 ˚C scan indicating that no WTe$_2$ or Te are present following growth at this temperature.

Some understanding of the influence of growth temperature can be achieved by examining the binary phase diagram for W-Te, reproduced in Fig. S3 from reference.(66) As the sample reaches the Te melting point - 450 ˚C - a liquid WTe$_x$ alloy forms at the boundary between the W and Te, as a result of the W dissolving into the liquid Te. It is clear from the phase diagram, that in this temperature range (450 ˚C to 650 ˚C) the alloy remains liquid for only a very small percentage W, beyond which it becomes supersaturated. This means that, as more W dissolves into the Te, solid WTe$_2$ crystals begin to nucleate and grow within the liquid alloy. Concurrent with this the Te is evaporating, further driving the growth of WTe$_2$ crystals. If the growth time is sufficiently long, all unreacted Te will evaporate, leaving a film solely consisting of WTe$_2$ during the cool-down. Below 450 ˚C the formation of an intermixed layer is greatly reduced due to the Te not melting, dramatically hindering the growth, as shown by the poor WTe$_2$ Raman signal at 400 ˚C and the absence of any discernible WTe$_2$ Raman signal at 350 ˚C.

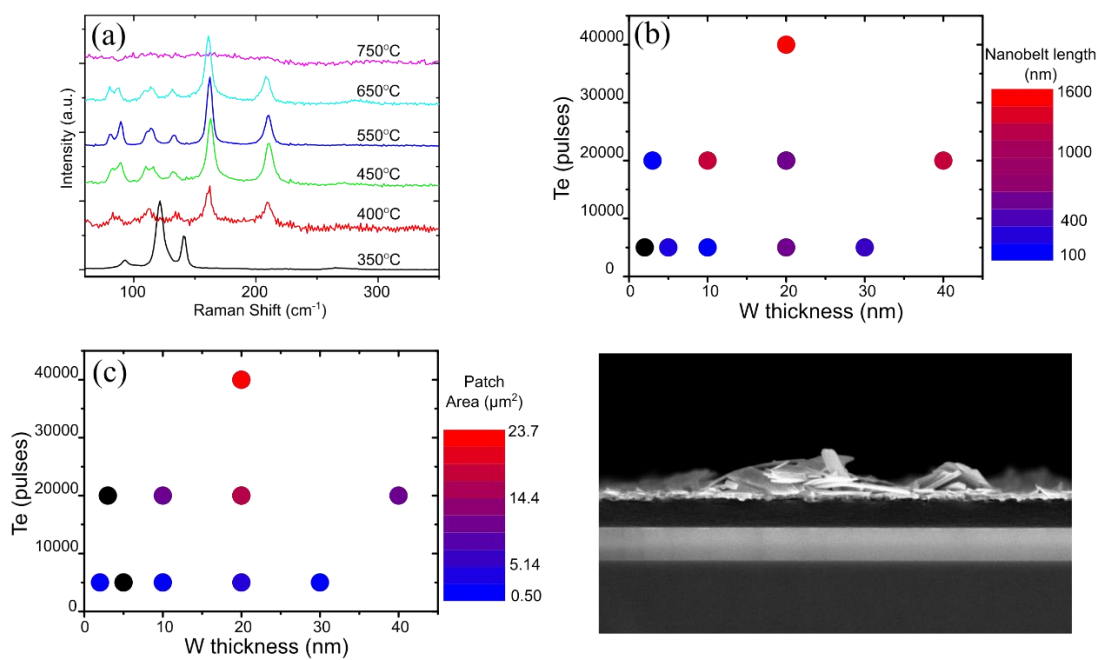

**Fig.3** (a), Raman of WTe2 films synthesised at a range of temperatures between 350 ˚C and 750 ˚C. At 350 ˚C only the signal for Te is present, between 450 ˚C and 650 ˚C there is a strong signal corresponding to bulk-like WTe$_2$. At 400 ˚C there is a weak WTe$_2$ signal and at 750 ˚C there are no discernible Raman peaks. (b) Plot of the variation average nanobelt length (100 nm – 1600 nm) with changes in the amount of W and Te present. The thickness of the initial layer of W was varied between 2 and 40 nm. Between 5,000 and 40,000 pulses of Te were electrodeposited on the surface. Red indicates longer nanobelt patches. (c) Variation of patch area with changes to W and Te amounts. (d) Cross-section SEM image of a cleaved substrate and WTe2 film showing raised patches of WTe$_2$ nanobelts.

The upper bound of ∼650 °C cannot be understood from the phase diagram, which indicates stability of WTe$_2$ up to ∼1000 °C. The proposed reason for the upper bound seen here is that the Te evaporates too quickly to react effectively, or any WTe$_2$ that does form decomposes through loss of Te due to the Te-deficient environment.(67, 68) This loss would not occur in CVT growth systems due to the vessel being sealed, maintaining high Te concentrations throughout the growth, allowing the use of higher growth temperatures.

By changing the initial quantity of W and Te on the samples it was possible to control the thickness of the films and also the average sizes of the WTe$_2$ grains/nanobelts. The average nanobelt size was primarily determined by the quantity of Te present, except for the thinnest films where both the W and Te quantity were important, as shown in Fig. 3(b). This figure shows W and Te quantities on the x- and y-axes respectively, with the average nanobelt size indicated by the colour of the data points. The lesser effect of W suggests that below a certain point it limits the growth, but above a particular threshold, it no longer has a strong influence on nanobelt size. The average length of the nanobelts could be varied between <100 nm and 1600 nm over the parameter space examined.

The size of the $WTe_2$ crystals grown in the film is affected by the Te quantity in two ways. Firstly, more Te provides more reactant in general, allowing larger crystal growth. Secondly, since the reaction to form $WTe_2$ occurs in the liquid Te alloy, more Te provides a greater duration for the reaction to occur before the Te vaporises.

The two distinct morphologies on the surface can be similarly explained. During the growth the Te dewets and forms droplets on the surface. These droplets serve as areas of high local reactant quantity, and long growth times, enabling much larger crystals to grow than in surrounding areas. Furthermore, the area of these patches containing larger $WTe_2$ nanobelts also increases with increasing Te quantity, as shown in Fig. 3(c).

These results correlate well with previously published reports for a similar system in the work of Kwak *et al*.(41) However a notable difference is that if one looks at a cross-section SEM of the films in this work, shown in Fig. 3(d), it is clear that the droplets were raised above the majority of the surface and that the $WTe_2$ crystals grew within them. This suggests that the W tends to dissolve into the Te rather than the Te etching down into the W, as was proposed in the previous work.

**2.4 Characterising Individual Nanobelts**

The majority of film conversion growth methods result in polycrystalline films with grain sizes between 10 and 100 nm. In contrast, the films produced in this work can be optimised to have nanobelts with lengths of up to 10 μm. The larger crystal size and discrete and separate nature of the nanobelts allowed us to develop a method to isolate individual $WTe_2$ nanobelts for further study by transferring them onto another substrate. This allowed examination of individual nanobelts in a manner not possible for a film. Details of this transfer method are given in Fig. S4.

Fig. 4(a) and (b) show optical and atomic force microscopy (AFM) images of the nanobelts on the destination substrate. The largest nanobelts measure about 10 μm long, width of ~600 nm and thicknesses of ~130 nm. While flakes with smaller lateral sizes were correspondingly thinner (10 – 20 nm), all nanobelts examined are nevertheless well into the range of displaying bulk-like characteristics.

The transfer of individual $WTe_2$ nanobelts made it possible to examine them using transmission electron microscopy (TEM). Initial analysis of the nanobelts was carried out using high-resolution TEM (HRTEM). Fig. 4(c) is a HRTEM image of a nanobelt, with the corresponding fast Fourier transform (FFT) shown in the insert. These highlight the crystalline nature of the $WTe_2$ with clear lattice fringes visible in the TEM image and discrete spots in the FFT.

To further examine the crystal structure and quality, annular dark-field scanning TEM (ADF-STEM) was carried out on the nanobelts. This analysis allowed atomic-resolution images to be captured, shown in Fig. 4(d). This image clearly shows that these nanobelts have a high level of crystallinity, with no grain boundaries visible over the area examined.

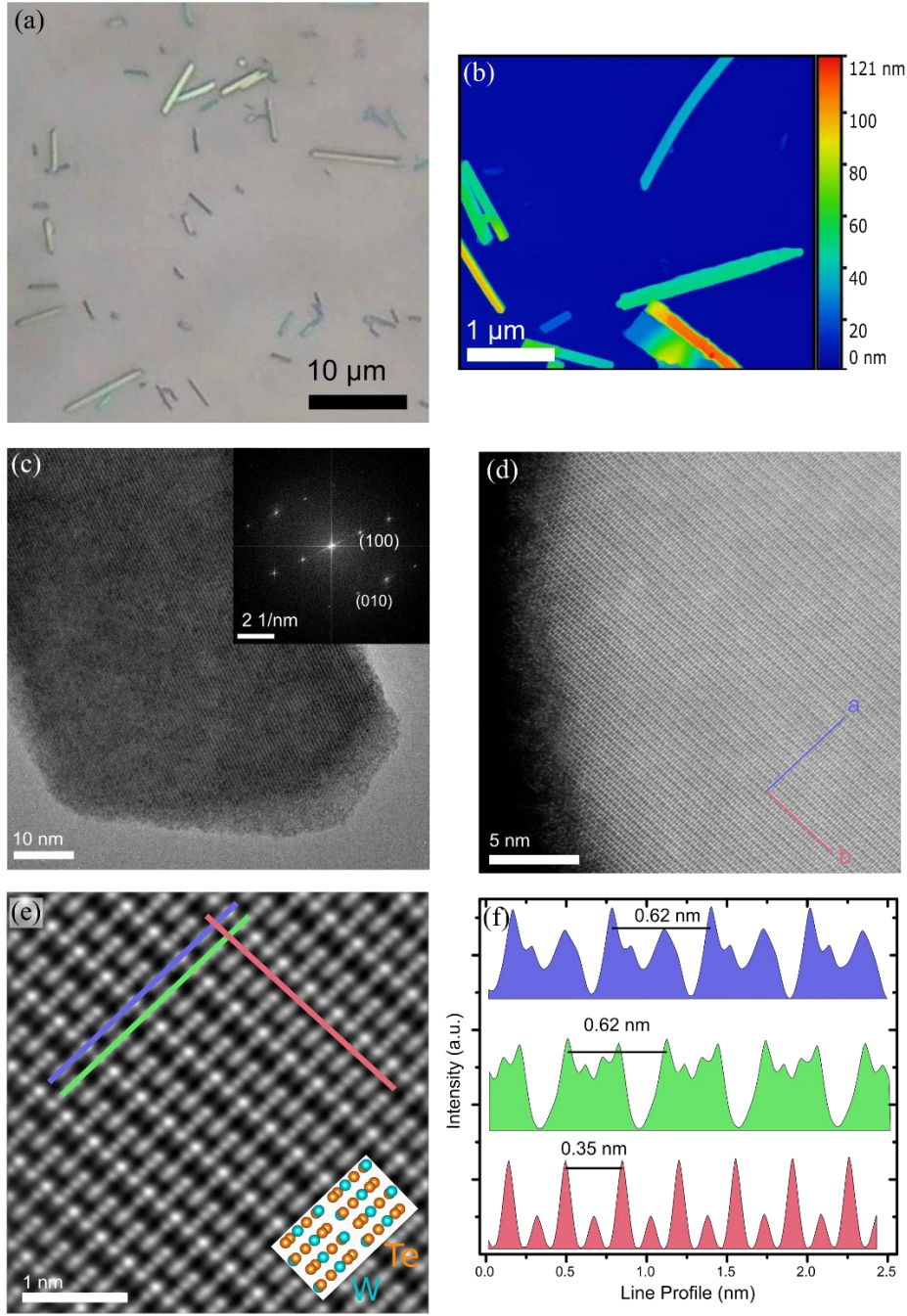

**Fig.4** (a) Optical image of transferred WTe$_2$ nanobelts. (b) AFM image of same sample (c) HRTEM image of a WTe$_2$ nanobelt, insert FFT of image (c). (d) ADF-STEM image of the edge of a WTe$_2$ nanobelt, with the WTe$_2$ crystallographic directions marked. (e) Atomic-resolution ADF image of WTe$_2$ crystal, compiled by averaging areas of (c). Insert: atomic positions of WTe$_2$ for appropriate crystallographic direction (f) Profiles of intensity measured from lines indicated in (d)

STEM images were taken from either end of a WTe$_2$ nanobelt, several micrometres in length, the FFTs of these images are shown in Fig. S5. These match to a very close degree, offering further confirmation of the single-crystalline nature of the nanobelts. It should be noted that there is some distortion in the images shown here. The orthogonal crystalline directions have measured angles of ~87 ° between them. This is thought to be most likely due to small stage drift or slight misalignment of the crystalline axis during imaging.(69)

The STEM analysis indicates the highly-crystalline nature of the nanobelts, without significant defect densities visible. As there are no visible grain boundaries within the nanobelts they are potentially comparable to mechanically-exfoliated flakes from CVT-grown crystals. This could open up a much quicker and lower-temperature synthesis route to produce micrometre size WTe$_2$ flakes for a range of studies.

By applying a spatial averaging and scan-distortion correction technique to Fig. 4(d), the image shown in Fig. 4(e) was obtained.(70) This averaging technique is discussed further in the methods secion.(71) This shows very clear atomic resolution giving the expected T$_d$ structure of WTe$_2$. This matches well to previously published analysis of WTe$_2$.(4, 41)

Measuring the intensity along the two crystallographic directions of this image, gave the line profiles shown Fig. 4(f).(72) The red profile shows a repeating pattern every 0.35 nm. This matches closely to the lattice constant of WTe$_2$ along the [010] crystallographic direction of 0.3496 nm. While the repeat period in the orthogonal direction of 0.62 nm, is very close to the unit cell length of 0.6282 nm for WTe$_2$ along the [100] direction. Using this information, along with the spacings obtained from the FFTs, allowed the conclusion to be drawn that it is the (001) lattice planes that are being examined in this case. A model of the expected atomic positions in this case is shown in Fig. 4(e) insert. This matches quite well to the alternating lines of atoms observed in the STEM image. When viewed from this direction, atoms in WTe$_2$ do not line up in perfect atomic columns meaning that exact atomic positions cannot be assigned despite the clarity of the image.

Examining again Fig. 4(c), it is possible to assign the lattice directions a and b, as shown by the arrows. This confirms that the W zigzag or 1D chains, which form along the b crystallographic direction in T$_d$ WTe$_2$, are parallel to the long axes of the nanobelt as expected.

## 2.5 Electrical characterisation

Electron-beam lithography (EBL) was used to contact individual WTe$_2$ nanobelts. This allowed the electrical behaviour of the crystalline nanobelts WTe$_2$ to be probed without the influence of junctions between WTe$_2$ grains. An AFM map of a device is shown in Fig. 5(a), with corresponding height profile in Fig. 5(b).

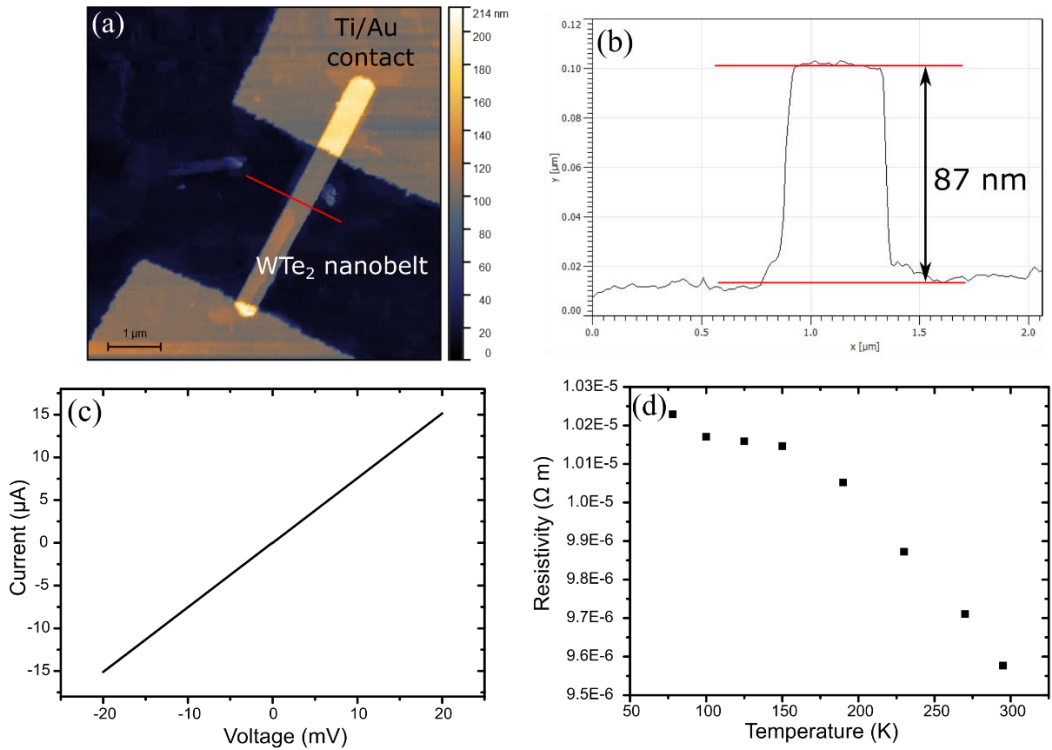

**Fig.5** (a) AFM image of the contacted WTe$_2$ nanobelt device (b) Height profile of the nanobelt along the line marked in (a). Two-terminal IV measurement of the WTe$_2$ nanobelt device. (d) Plot showing the change in resistivity of the nanobelt as the temperature is increased from 78K to 293K

The two-terminal IV response of a nanobelt device is shown in Fig. 5(c). The device response is Ohmic over the measured region. The dimensions from AFM allowed the resistivity of the device to be calculated, this was found to be 9.3 x $10^{-6}$ Ωm. This low resistivity is expected due to the semimetallic nature of WTe$_2$. Interestingly, this resistivity matches almost exactly the values measured by Song *et al.* in their paper on the potential use of single-crystal WTe$_2$ nanobelts as electrical interconnects in nanoelectronics.(23) Their synthesis method requires the deposition of films of W and Cu that are then exposed to Te vapour. This causes a CuTe$_x$ alloy to form, in which the WTe$_2$ grows. Following this a post-growth etch is needed to remove residual CuTe$_x$.(41) The synthesis method in this work is more straight-forward and does not require a post-growth etch offering a potential advantage. The nanobelts produced here have a very similar form, crystallinity and electrical response to the work of Song *et al.*, indicating that the WTe$_2$ is likely of a similar quality and thus suitable for similar applications.

The temperature dependence of the WTe$_2$ nanobelt resistance was investigated, with the results shown in Fig. 5(d). The nanobelt shows very little change in resistivity with temperature; becoming ~10% less resistive as the temperature is increased from 77K to 300K. A small change in resistivity is unsurprising for a semimetallic material, though a trend of decreasing resistivity with temperature is more usually associated with a semiconductor. There is some variation in literature reports of WTe$_2$ resistivity change as a function of temperature. Some works on thin crystals of WTe$_2$ show a reduction in resistivity with increasing temperature due to Anderson localization effects. However that was for much thinner (3 - 4 layer) flakes, while the nanobelt examined here was ~120 layers.(15, 23) Works on thicker samples of WTe$_2$ generally show increasing resistivity as the temperature is increased. The magnitude of this effect is reported as being between ~5% and ~500% change over a similar temperature window to that examined in this work.(23, 32, 54, 73) It is possible that effects such as the formation of a potential barrier at the contacts due to surface oxidation may also be influencing results seen in this and other works. Further studies would be required to fully understand the observed result.

## 3. CONCLUSIONS

We developed a method to synthesise large-area films of WTe$_2$ from pre-deposited W and Te layers. Successful conversion of these layers to WTe$_2$ films was confirmed with various spectroscopies. SEM showed that the polycrystalline films consisted of two different morphologies, one being regions of nanobelts up to 10 μm in length. The films were found to be stoichiometric WTe$_2$ but with a significant amount of surface oxidation. In line with previous reports of WTe₂ oxidation being self-limiting, the oxide level in the bulk of the film was found to be substantially lower.

Further to this we isolated individual nanobelts from the films to allow additional characterisation. Through TEM analysis the nanobelts were shown to be highly crystalline with low defect densities. Combined with promising initial electrical results, this indicates the potential for these nanobelts to be used in investigations where previously mechanically-exfoliated CVT crystals were used.

This ability to yield both WTe₂ films and nanobelts in a straight-forward manner at relatively low synthesis temperature is a compelling advantage of this growth method over those used in many previous studies.

## 4. EXPERIMENTAL METHODS

A film of W was deposited onto the substrate using a Gatan PECS Ar-ion sputtering tool. A layer of Te was subsequently deposited on top of this by electrodeposition. Te was reduced from solution, 0.02 M TeO₂ in 1 M nitric acid, in an electrochemical cell. A platinum counter electrode and a Ag/AgCl reference electrode were used. This is shown in Fig. S1(b). The reduction proceeded *via* the reaction(74):

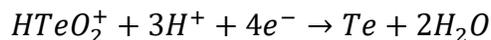
$$HTeO_2^+ + 3H^+ + 4e^- \rightarrow Te + 2H_2O$$

A pulsing sequence was used during the deposition, this consisted of 10 ms pulses of -0.3 V applied to the working electrode with respect to the reference electrode with 50 ms gaps between these. Varying the number of pulses in a deposition served to control the quantity of Te deposited on the sample. Using a pulsing potential was found to yield a more uniformly deposited film of Te.(74) Further detail is given in a previous work using a similar method to synthesise $MoTe_2$.(56)

The substrates used throughout the work are PyC on a 300 nm thermal $SiO_2$ on Si wafer. The PyC layer is not necessary but served to improved conductivity during the electrodeposition. The PyC was grown by CVD of acetylene at 950°C for 30 mins on $SiO_2$/Si substrates in a hot-wall, quartz-tube furnace.(75)

Following the Te deposition, the films were converted to $WTe_2$ in an ATV PEO 604 quartz furnace under a nitrogen atmosphere at a pressure of ~700 mbar. The growth temperature was typically 550 °C but other growth temperatures were also investigated. Nested crucibles were used to maintain a high partial pressure of Te in the vicinity of the samples while avoiding contaminating the furnace with high quantities of Te, as shown in Fig. S1(a). The temperature was ramped at 180 °C $min^{-1}$ and held at the growth temperature for 90 mins. The samples were then allowed to cool to near room temperature (<30 °C) under $N_2$ over a period of ~3 hours before removal from the furnace. During the cooling, at ~320 °C, the pressure was lowered from ~700 mbar to 13 mbar to ensure removal of any remaining Te. All samples are named by referring to the thickness of the starting W film.

Nanobelts of $WTe_2$ were mechanically transferred from the growth substrate using adhesive Nitto Denka BT-150E-CM tape. A schematic of this process is shown in Fig. S4. In order to grow the films with the largest possible $WTe_2$ nanobelts, samples with 40 nm of W and 40,000 pulses of Te were used for transferring nanobelts from.

SEM images were obtained with a Karl Zeiss Supra microscope operating at 3 kV accelerating voltage, 30 μm aperture and a working distance of ~3-4 mm.

The XRD measurement was performed on a Bruker D8 Discover with a monochromated Cu K-alpha source.

A WITec Alpha 300R with a 532 nm excitation laser, with a power of ~200 µW, was used to collect the Raman spectra shown herein. All Raman measurements were taken using a spectral grating with 1800 lines/mm and a 100x objective lens. Raman spectra shown are averages of maps which were generated by taking scans every 400 nm in the x and y directions, typically over an area of 20 x 20 µm, making each spectrum an average of ~2,500 spectra.

X-ray photoelectron spectroscopy (XPS) spectra were taken with a PHI VersaProbe III instrument equipped with a micro-focused, monochromatic Al Kα source (1486.6 eV) and a dual-beam charge neutralization was used. Core level spectra were recorded with a spot size of 100 µm and a pass energy of 69 eV using PHI SmartSoft VersaProbe software, and processed with PHI MultiPak 9.8. Sputter XPS depth profiling was conducted using 1 keV Ar ions. Binding energies were referenced to the adventitious carbon signal at 284.8 eV. Spectral components were fitted using a Shirley background subtraction and appropriate line shapes. Relative atomic percentages were calculated using the relative sensitivity factors provided by the software CasaXPS

AFM was carried out on a Bruker Multimode 8 in ScanAsyst Air mode using Nanosensor PointProbe Plus tips.

HRTEM analysis was performed in an FEI Titan TEM at an acceleration voltage of 300 kV. Atomic-resolution STEM images were obtained with a Nion UltraSTEM200, using a HAADF detector, operated at 200 kV.

An overview of how STEM averaging was performed is that initially a small area of the sample (4 nm x 4 nm in this case) with atomic resolution is chosen. The software then scans the remainder of the image and finds all the areas with which this image matches to a high degree. All of these areas (300 in this analysis) are then extracted and stacked to form an average image.(71) Further image processing then removes any distortion in the image.(70)

For electrical measurements samples were patterned by EBL before Ti/Au (5 nm/100 nm) contacts were deposited using electron-beam evaporation.

## ACKNOWLEDGEMENTS

J.B. Mc M. acknowledges an Irish Research Council scholarship, Project 204486, Award 13653. N. M. acknowledges support from SFI through 15/SIRG/3329 and 12/RC/2278_P2. G. S. D. acknowledges the support of SFI through PI_15/IA/3131. LJ acknowledges SFI and the Royal Society Fellowship URF/RI/191637. The SEM and (S)TEM imaging for this project was carried out at the Advanced Microscopy Laboratory (AML), Trinity College Dublin, Ireland. The AML (www.tcd.ie/crann/aml) is an SFI supported imaging and analysis centre, part of the CRANN Institute and affiliated to the AMBER centre.

## ETHICAL STANDARDS

The authors declare that the experiments in this work comply with the current lasws of the country in which they were performed.

## CONFLICT OF INTEREST

There are no conflicts of interest to declare.

# Synthesis of WTe$_2$ thin films and highly-crystalline nanobelts from pre-deposited reactants

*John B. Mc Manus[1,2], Cansu Ilhan[1,2], Bastien Balsamo[2,3], Clive Downing[1,2], Conor P. Cullen[1,2], Tanja Stimpfel-Lidner[4], Graeme Cunningham[1,2], Lisanne Peters[1,2], Lewys Jones[2,5], Daragh Mullarkey[5], Igor V. Shvets[5], Georg S. Duesberg[1,4], Niall McEvoy[1,2]\**

[1] School of Chemistry, Trinity College Dublin, Dublin 2, D02 PN40, Ireland

[2] AMBER Centre, CRANN Institute, Trinity College Dublin, Dublin 2, D02 PN40, Ireland

[3] SIGMA Clermont, Université Clermont Auvergne, F-63000 Clermont–Ferrand, France

[4] Institute of Physics, EIT 2, Faculty of Electrical Engineering and Information Technology, Universität der Bundeswehr, 85579 Neubiberg, Germany

[5] School of Physics, Trinity College Dublin, Dublin 2, D02 PN40, Ireland

\*Corresponding author: nmcevoy@tcd.ie

# Supporting Information

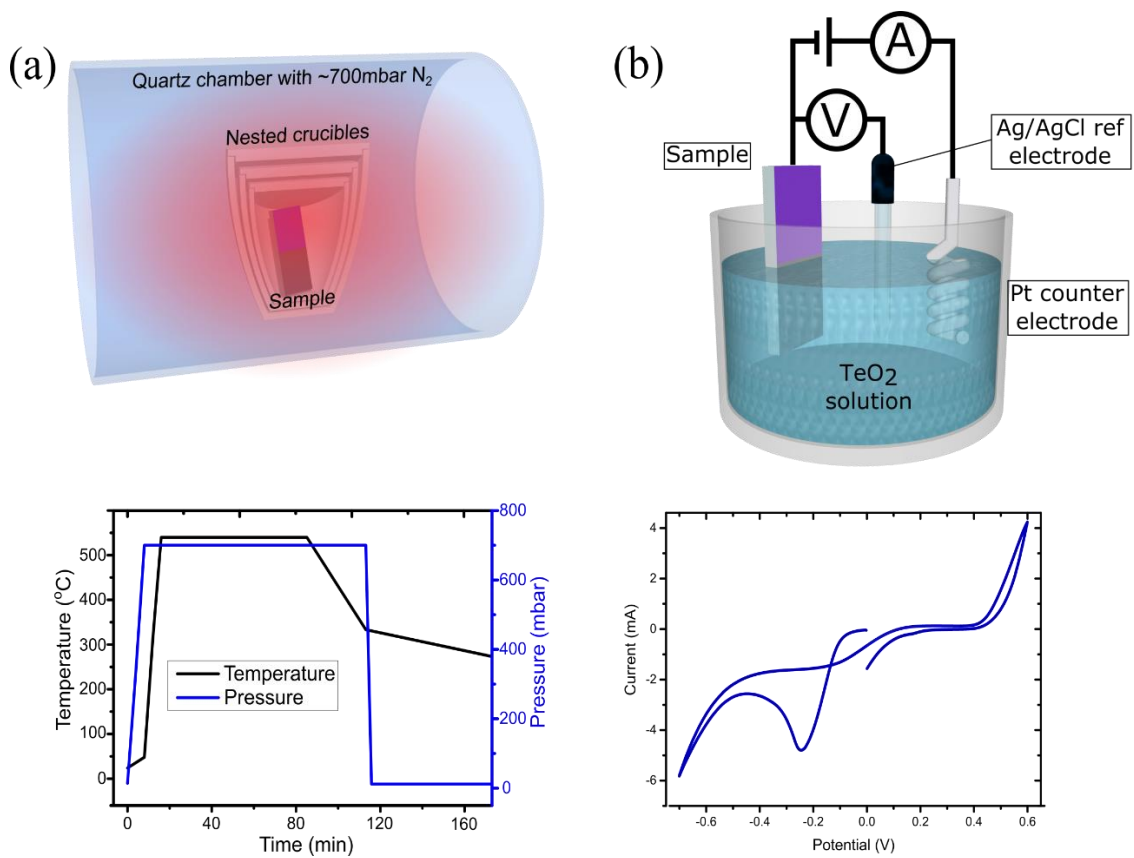

Fig S1: (a) Diagram of the quartz furnace and nested crucibles used to synthesize the $WTe_2$ films and nanobelts. Illustrative plot of typical temperature and pressure conditions in the furnace during the growth of the $WTe_2$. (b) Schematic of the Te electrodeposition process with the sample as the working electrode. Cyclic voltammagram of the cell showing the reduction peak at ~ -0.2V corresponding to the electrodeposition of Te on the sample.

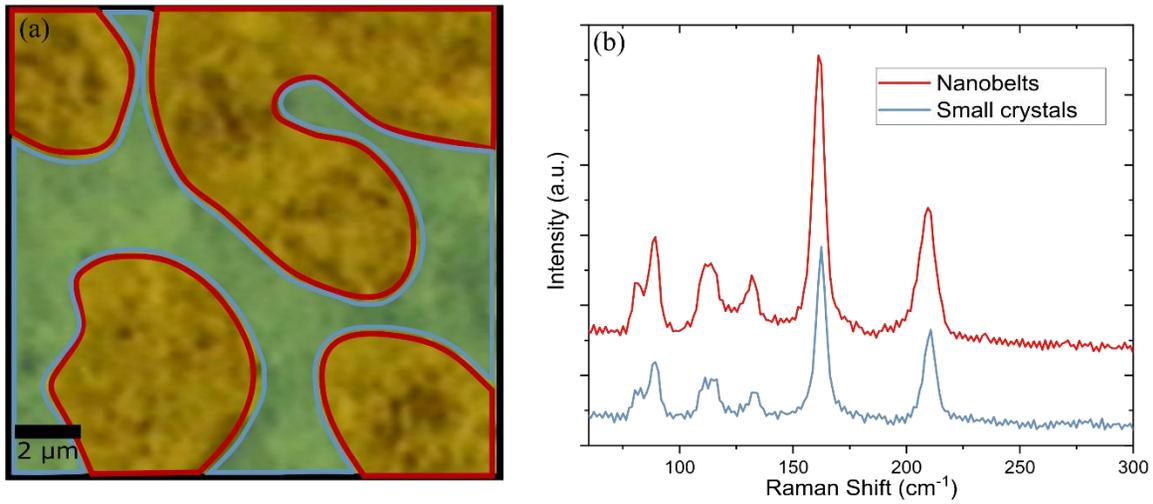

Fig. S2 (a) Optical image of a $WTe_2$ film. The areas enclosed in red are the patches of nanobelts. The areas highlighted blue are the smaller crystals that surround the patches of nanobelts. (b) Average Raman spectra corresponding to the two different morphology areas shown in (a). The red trace is the average Raman spectrum of the patches of nanobelts – red area in (a). The blue spectrum corresponds to the areas of smaller crystals. The expected $WTe_2$ modes in both indicates the presence of $WTe_2$ in both areas. The blue spectrum is less intense and slightly noisier, likely due to the smaller crystallite size in this area.

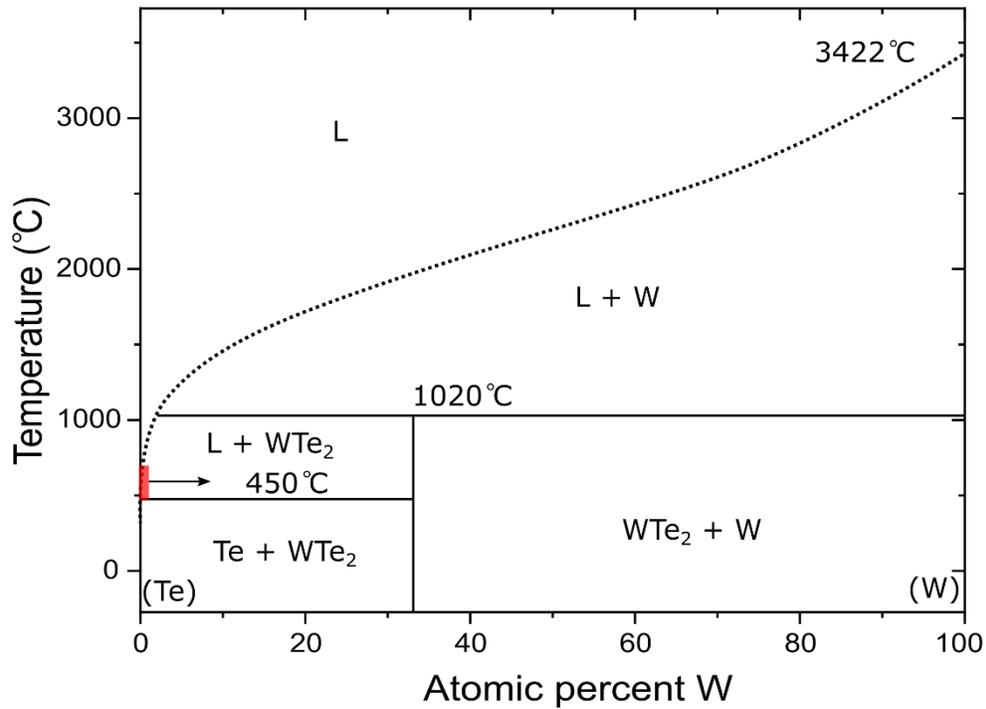

Fig. S3 Binary phase diagram for W-Te with temperature shown in degree Celcius, reproduced from reference.[1] The red area indicates the initial growth conditions for the WTe$_2$ in the liquid WTe$_x$ alloy. The arrow indicates the direction that the overall system moves as the WTe$_2$ forms and the remaining Te evaporates.

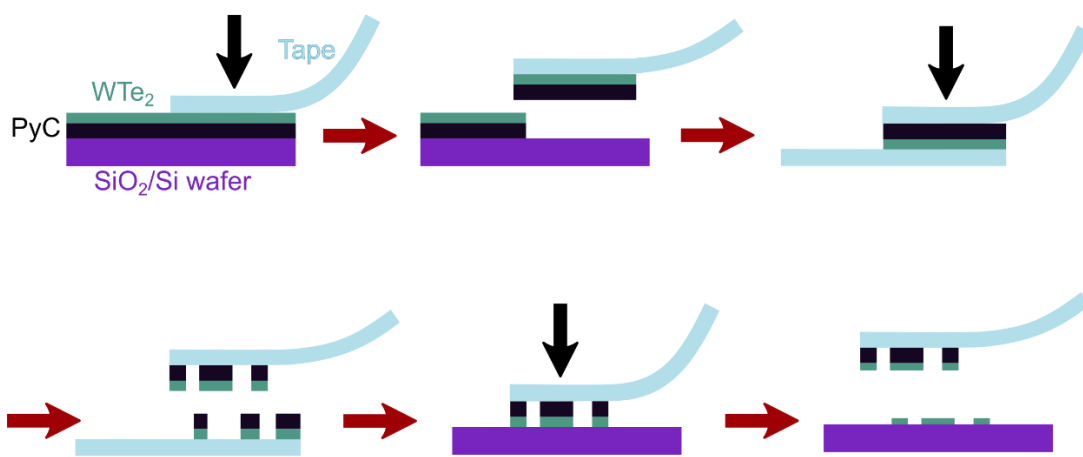

Figure S4: All samples were grown on PyC to aid the Te electrodeposition, however this did complicate the transfer somewhat. Firstly, a piece of tape was pressed down onto the sample with appropriate force (firm pressure under finger), before being removed, taking the $WTe_2$ and PyC layers with it. A second piece of tape was then pressed onto the first and removed. This second piece then had PyC in contact with the tape, and some $WTe_2$ nanobelts on top of this. This piece was then pressed onto a fresh $SiO_2$/Si wafer substrate and removed, resulting in $WTe_2$ nanobelts being deposited. This piece of tape could be used to deposit on multiple substrates, with each repetition giving less dense arrays of $WTe_2$ nanobelt. This technique was successfully used to transfer nanobelts for further analysis by TEM and for electrical device fabrication.

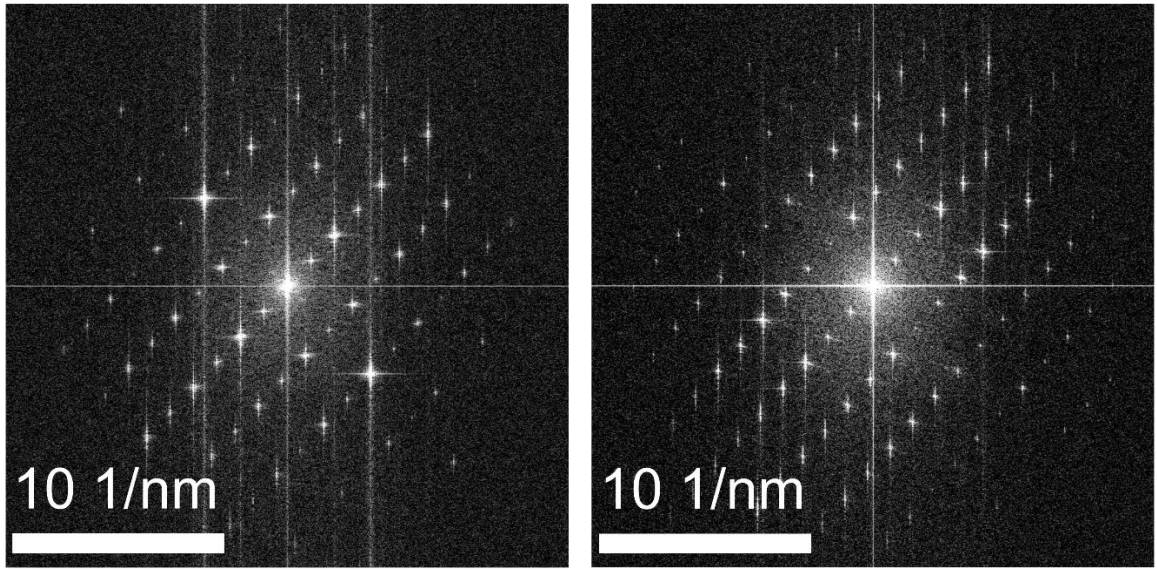

Fig. S5 FFT of images of opposite ends of a WTe$_2$ nanobelt. The similarity of these indicates that the same crystalline phase and orientation is present through the nanobelt.

References
1. Predel, B., *Phase Equilibria, Crystallographic and Thermodynamic Data of Binary Alloys*. Springer: Vol. 5.